\documentclass[12pt,a4paper]{article}

\usepackage{amsmath}
\usepackage{amssymb}
\usepackage{epsfig}
\usepackage{graphicx}
\usepackage{cite}

\newcommand{\capdef}{}
\newcommand{\mycaption}[2][\capdef]{\renewcommand{\capdef}{#2}%
       \caption[#1]{{\footnotesize #2}}}
\makeatletter
\renewcommand{\fnum@table}{\textbf{\tablename~\thetable}}
\renewcommand{\fnum@figure}{\textbf{\figurename~\thefigure}}
\makeatother

\newcounter{myenumi}

\renewcommand{\themyenumi}{\roman{myenumi}}
{\end{list}}

\newlength{\myem}
\newcounter{mysubequation}[equation]

\makeatletter
\renewcommand{\section}{\@startsection{section}{1}{0em}{-\baselineskip}%
{\baselineskip}{\normalfont\large\bfseries}}
\renewcommand{\subsection}%
{\@startsection{subsection}{2}{0em}{-0.7\baselineskip}%
{0.7\baselineskip}{\normalfont\bfseries}}
\makeatother

\newcommand{\ie}{{\it i.e.}}

\newcommand{\eg}{{\it e.g.}}

\newcommand{\cf}{{\it cf.}}

\newcommand{\eq}{Eq.}

\newcommand{\fig}{Fig.}

\newcommand{\Ref}{Ref.}
\newcommand{\Refs}{Refs.}

\newcommand{\bi}{\begin{itemize}}
\newcommand{\ei}{\end{itemize}}

\newcommand{\be}{\begin{equation}}
\newcommand{\ee}{\end{equation}}
\newcommand{\bea}{\begin{eqnarray}}
\newcommand{\eea}{\end{eqnarray}}

\newcommand{\sdm}{\Delta m_{21}^2}
\newcommand{\deltacp}{\delta_{\mathrm{CP}}}
\newcommand{\stheta}{\sin^2 2 \theta_{13}}

\newcommand{\MINOS}{{\sf MINOS}}
\newcommand{\ICARUS}{{\sf ICARUS}}
\newcommand{\OPERA}{{\sf OPERA}}
\newcommand{\JHFSK}{{\sf JHF-SK}}
\newcommand{\NUMI }{{\sf NuMI}}

\newcommand{\JHFHK}{{\sf JHF-HK}}

\newcommand{\GLOBES}{{\sf GLoBES}}
\newcommand{\AEDL}{{\sf AEDL}}

\newcommand{\equ}[1]{\eq~(\ref{equ:#1})}
\newcommand{\figu}[1]{\fig~\ref{fig:#1}}

\begin{document}

\begin{titlepage}

\renewcommand{\thefootnote}{\alph{footnote}}

\vspace*{-3.cm}
\begin{flushright}
TUM-HEP-553/04\\
\end{flushright}

\vspace*{0.5cm}

\renewcommand{\thefootnote}{\fnsymbol{footnote}}
\setcounter{footnote}{-1}

{\begin{center}
{\Large\bf Simulation of long-baseline neutrino oscillation
experiments with GLoBES}
\end{center}}
{\begin{center}
{\large\bf (General Long Baseline Experiment Simulator)}
\end{center}}
\renewcommand{\thefootnote}{\alph{footnote}}

\vspace*{.8cm}
{\begin{center} {\large{\sc
                P.~Huber\footnote[1]{\makebox[1.cm]{Email:}
                phuber@ph.tum.de},~
                M.~Lindner\footnote[2]{\makebox[1.cm]{Email:}
                lindner@ph.tum.de},~and~
                W.~Winter\footnote[3]{\makebox[1.cm]{Email:}
                wwinter@ph.tum.de}
                }}
\end{center}}
\vspace*{0cm}
{\it
\begin{center}

\footnotemark[1]${}^,$\footnotemark[2]${}^,$\footnotemark[3]%
       Institut f\"ur Theoretische Physik, Physik--Department,\\
       Technische Universit\"at M\"unchen,
       James--Franck--Strasse, D--85748 Garching, Germany

\end{center}}

\vspace*{1cm}

\begin{abstract}
We present the \GLOBES\ (``General Long Baseline Experiment Simulator'')
software package, which allows the simulation of long-baseline and reactor
neutrino oscillation experiments. One part of the software is the
abstract experiment definition language to define experiments with
beam and full detector descriptions as accurate as possible. Many systematics options are provided, such as normalization and energy
calibration errors, or the choice between spectral or total rate 
information. For the 
definition of experiments, a new transparent building block concept is 
introduced. In addition, an additional program provides the possibility 
to develop and test new experiment definitions quickly.
Another
part of \GLOBES\ is the user's interface, which provides probability, rate, 
and $\Delta \chi^2$ information for a given experiment or any
combination of up to 32 experiments in C. Especially, the
$\Delta \chi^2$ functions allow a simulation with statistics only,
systematics, correlations, and degeneracies. In particular, \GLOBES\
can handle the full multi-parameter correlation among the oscillation
parameters, external input, and matter density uncertainties. 
\end{abstract}

\vspace*{.5cm}

\end{titlepage}

\newpage

\renewcommand{\thefootnote}{\arabic{footnote}}
\setcounter{footnote}{0}


\section{Introduction}

Neutrino oscillations are now established as the leading flavor
transition mechanism for neutrinos. In a long history of many experiments, see
\eg~\cite{Barger:2003qi}, 
two oscillation
frequencies have been identified: The fast ``atmospheric'' and the
slow ``solar'' oscillations, which are driven by the respective
mass squared differences.  In addition, there could be interference
effects between these two oscillations, provided that the coupling given by 
the small mixing angle $\theta_{13}$ is large enough. Such interference
effects include, for example, leptonic CP violation.  In order to
test the unknown oscillation parameters, \ie, the mixing angle $\theta_{13}$,
 the leptonic CP phase, and the
neutrino mass hierarchy, new long-baseline and reactor experiments 
are proposed.
These experiments send an artificial neutrino beam to a detector, or
detect the neutrinos produced by a nuclear fission reactor. However,
the presence of multiple solutions which are intrinsic to
 neutrino oscillation probabilities~\cite{Fogli:1996pv,
Burguet-Castell:2001ez,Minakata:2001qm,Barger:2001yr} 
affect these measurements.
Thus optimization strategies are required which maximally exploit 
complementarity between experiments. Therefore, a modern, complete experiment 
simulation  and analysis tool does not only need to have 
a highly accurate beam and detector simulation, but also powerful 
means to analyze correlations and degeneracies, especially for the combination
of several experiments.
In this paper, we present the \GLOBES\ software package
as such a tool.

\begin{figure}[t]
\begin{center}
\includegraphics[width=14cm]{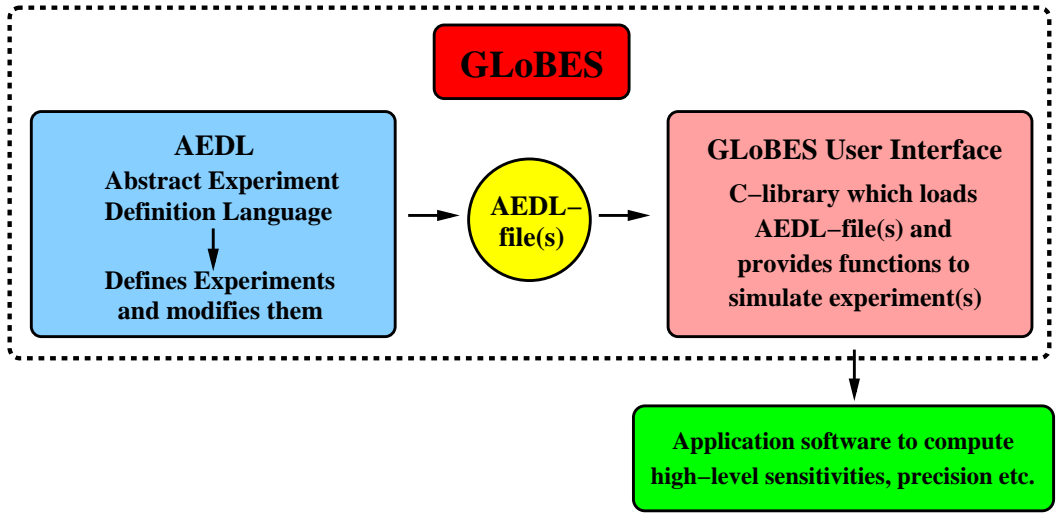}
\end{center}
\caption{\label{fig:GLOBES} General concept of the \GLOBES\ package.}
\end{figure}

\GLOBES\ (``General Long Baseline Experiment Simulator'') is a flexible
software tool to simulate and analyze neutrino oscillation 
long-baseline and reactor experiments using a 
complete three-flavor description. On the
one hand, it contains a comprehensive abstract experiment definition
language (AEDL), which allows to describe 
most classes of long baseline and reactor experiments
at an abstract level. On the other hand, it provides a C-library to 
process the experiment information in order to obtain oscillation
probabilities, rate vectors, and $\Delta \chi^2$-values (\cf, \figu{GLOBES}). 
In addition, it provides a binary program to test experiment
definitions very quickly, before they are used by the application software.
Currently, \GLOBES\ is available for GNU/Linux. Since the source code is 
included, the modifications to other operating systems should be doable.

\GLOBES\ allows to simulate experiments with stationary neutrino point 
sources, where each experiment is assumed to have only one neutrino source.
Such experiments are neutrino beam experiments and reactor experiments. 
Geometrical effects of a source distribution, such as in the sun or the 
atmosphere, can not be described. In addition, sources with a physically 
significant time dependencies  can not be studied, such as  supernov\ae. 

On the experiment definition side, either built-in neutrino fluxes
(\eg, neutrino factory) or arbitrary, user defined fluxes can be used. 
Similarly,
arbitrary cross sections, energy dependent efficiencies,
energy resolution functions as well as the considered oscillation channels, 
backgrounds, and many other properties can be specified. 
For the systematics, energy
normalization and calibration errors can be simulated. Note that
energy ranges and windows and bin widths can be
(almost) arbitrarily chosen, including bins of different
widths. Together with \GLOBES\ comes a number of
pre-defined experiments in order to demonstrate the capabilities
of \GLOBES\ and to provide prototypes for new experiments.
In addition, they can be used to test new physics ideas with
complete experiment simulations.
Examples for these prototypes are the \MINOS , \ICARUS , and
\OPERA\ simulations from \Ref~\cite{Huber:2004ug}, the
\JHFSK\ and \NUMI\ superbeam simulations from \Refs~\cite{Huber:2002mx,
Huber:2002rs}, the \JHFHK\ superbeam upgrade simulation from
\Ref~\cite{Huber:2002mx}, the neutrino factory simulations from
\Ref~\cite{Huber:2002mx}, and the reactor experiment simulations from
\Ref~\cite{Huber:2003pm}. Other projects using earlier versions
of the \GLOBES\ software include \Refs~\cite{Huber:2003ak,Ohlsson:2003ip,Winter:2003ye,Antusch:2004yx}.

With the C-library, one can extract the $\Delta \chi^2$ for all defined 
oscillation channels for an experiment or any combination of experiments.
Of course, also low-level information, such as oscillation
probabilities or event rates, can be obtained. \GLOBES\ includes the
simulation of neutrino oscillations in matter with arbitrary matter 
density profiles, as well as it allows to simulate the matter density
uncertainty. As one of the most
advanced features of \GLOBES , it provides the technology to 
project the $\Delta \chi^2$, which is a function of all oscillation
parameters, onto any subspace of parameters by local minimization. 
This approach allows the inclusion of multi-parameter-correlations,
where external constraints (\eg, on the solar parameters) can be imposed, too.
Applications of the projection mechanism include the projections onto 
the $\stheta$-axis and the $\stheta$-$\deltacp$-plane. In addition, 
all oscillation parameters can be kept free to precisely localize 
degenerate solutions.

\section{The computation of raw event rates}

In this section, we sketch the computation of 
the event rates, which is one of the core parts of the \GLOBES\
software. However, because of the complexity of this issue,
we refer to the \GLOBES\ manual~\cite{Manual} for
details.  

The differential event rate for each channel is given by
\begin{eqnarray}
\label{equ:master_event}
\frac{dn_{\beta}^{\text{IT}}}{dE'}=&&N\,\int\limits_0^\infty \int\limits_0^\infty dE\,d\hat{E}\quad
\underbrace{\Phi_{\alpha} (E)}_{\mathrm{Production}} \times \nonumber\\
&&\underbrace{\frac{1}{L^2} P_{(\alpha\rightarrow\beta)}(E,L,\rho;\theta_{23},
\theta_{12},\theta_{13},
\Delta m^2_{31},\Delta m^2_{21},\deltacp)}_{\mathrm{Propagation}}
\times \nonumber \\ &&\underbrace{\sigma^{\text{IT}}_f(E)
k_f^{\text{IT}}(E-\hat{E})}_{\mathrm{Interaction}} \times \nonumber \\
&&\underbrace{ T_f(\hat{E}) V_f(\hat{E}-E')}_{\mathrm{Detection}}\,,
\end{eqnarray}
where $\alpha$ is the initial flavor of the neutrino, 
$\beta$ is the final flavor, $\Phi_{\alpha} (E)$ is the flux of the 
initial flavor at the
source, $L$ is the baseline length, $N$ is a normalization factor, and 
$\rho$ is the matter density. The energies in this formula are given as follows:
\begin{itemize}
\item
 $E$ is the incident neutrino energy, \ie, the actual energy of the 
incoming neutrino (which is not directly accessible to the experiment)
\item
 $\hat{E}$ is the energy of the secondary particle
\item
 $E'$ is the reconstructed neutrino energy, \ie, the measured
neutrino energy as obtained from the experiment
\end{itemize}
The interaction term is composed of 
two factors, which are the total cross section 
$\sigma^{\text{IT}}_\beta(E)$ for the flavor $f$ and
the interaction type IT, and the energy distribution of the 
secondary particle $k_\beta^{\text{IT}}(E-\hat{E})$.
The detector properties are 
modeled by the threshold function $T_\beta(\hat{E})$, coming from the the 
limited resolution or the cuts in the analysis, and the energy resolution 
function $V_\beta(\hat{E}-E')$ of the secondary particle. 

Since it is a lot of effort to solve this double integral numerically,
we split up the two integrations. First, we evaluate the integral over
$\hat{E}$, where the only terms containing $\hat{E}$ are
$k_\beta^{\text{IT}}(E-\hat{E})$,  $ T_\beta(\hat{E})$, and 
$ V_\beta(\hat{E}-E')$. We define:
\begin{eqnarray}
\label{equ:e_res} 
R_\beta^{\text{IT}}(E,E')\,\epsilon_\beta^{\text{IT}}(E')
 \equiv
\int\limits_0^\infty d\hat{E} \quad T_\beta(\hat{E})\,k_\beta^{\text{IT}}(E-\hat{E})
\,V_\beta(\hat{E}-E')\,. 
\end{eqnarray}
Thus, $R_\beta^{\text{IT}}(E,E')$ describes the energy response of 
the detector, \ie , a neutrino with a (true) energy $E$ is reconstructed
with an energy between $E'$ and $E'+dE'$ with a probability
$R_\beta^{\text{IT}}(E,E') dE'$. The function $R(E,E')$ is also often called ``energy resolution function''. Actually, its internal representation
in the software is a smearing matrix. The function $\epsilon_\beta^{\text{IT}}(E')$ will be refered to as ``post-smearing efficiencies'', since it will allow us to define cuts and threshold functions {\em after} the smearing is performed, \ie, as function of $E'$. 
In addition, \GLOBES\ uses ``pre-smearing efficiencies'', which are 
evaluated {\em before} the smearing is performed, \ie, as function of $E$.
Similarly, constant\footnote{With respect to the oscillation parameters, 
not the energy.} background rates can be added to the event rates before or 
after the energy smearing, which are refered to as 
``pre-smearing backgrounds'' and ``post-smearing backgrounds''. These types of 
efficiencies and backgrounds allow a very accurate modeling of
many factors, such as atmospheric or geoneutrino backgrounds, energy cuts, or
energy threshold functions.

Eventually, we can write down the number of events per bin $i$ \index{Bin} and channel $c$ as
\begin{equation}
\label{equ:channel}
n_i^c=\int_{E_i-\Delta E_i/2}^{E_i+\Delta E_i/2} dE' \quad
\frac{dn_{\beta}^{\text{IT}}}{dE'} (E') \,
\end{equation}
where $\Delta E_i$ is the bin size of the $i$th energy bin.
This means that one has to solve the integral
\begin{eqnarray}
\label{equ:events_bin}
n_i^c=N/L^2\,\int_{E_i-\Delta E_i/2}^{E_i+\Delta E_i/2} dE' 
\quad \int\limits_0^\infty dE \,\, \Phi^c(E)\,
P^c(E)\,
\sigma^c(E)\,
R^c(E,E')\,
\epsilon^c(E')\,,
\end{eqnarray} 
which gives the raw event rates of the channel $c$ in the $i$th bin.
Note that the events are binned according to their \emph{reconstructed} energy.

Core part of the event rate computation is the energy smearing algorithm
to evaluate \equ{events_bin}, where either a particular energy resolution
 function can be chosen for automated energy smearing, or the discretized
smearing matrix $R^c(E,E') = R^c_{ij} $ can be given manually. In addition, it is possible
to define a low-pass filter to avoid aliasing effects for neutrino
oscillations faster than the sampling width.

\section{Definition of experiments with \AEDL\\ (Abstract Experiment 
Definition Language)}

\begin{figure}[t]
\begin{center}
\begin{tabular}{p{6.3cm}p{4.3cm}p{4.3cm}}
\includegraphics[width=6.5cm]{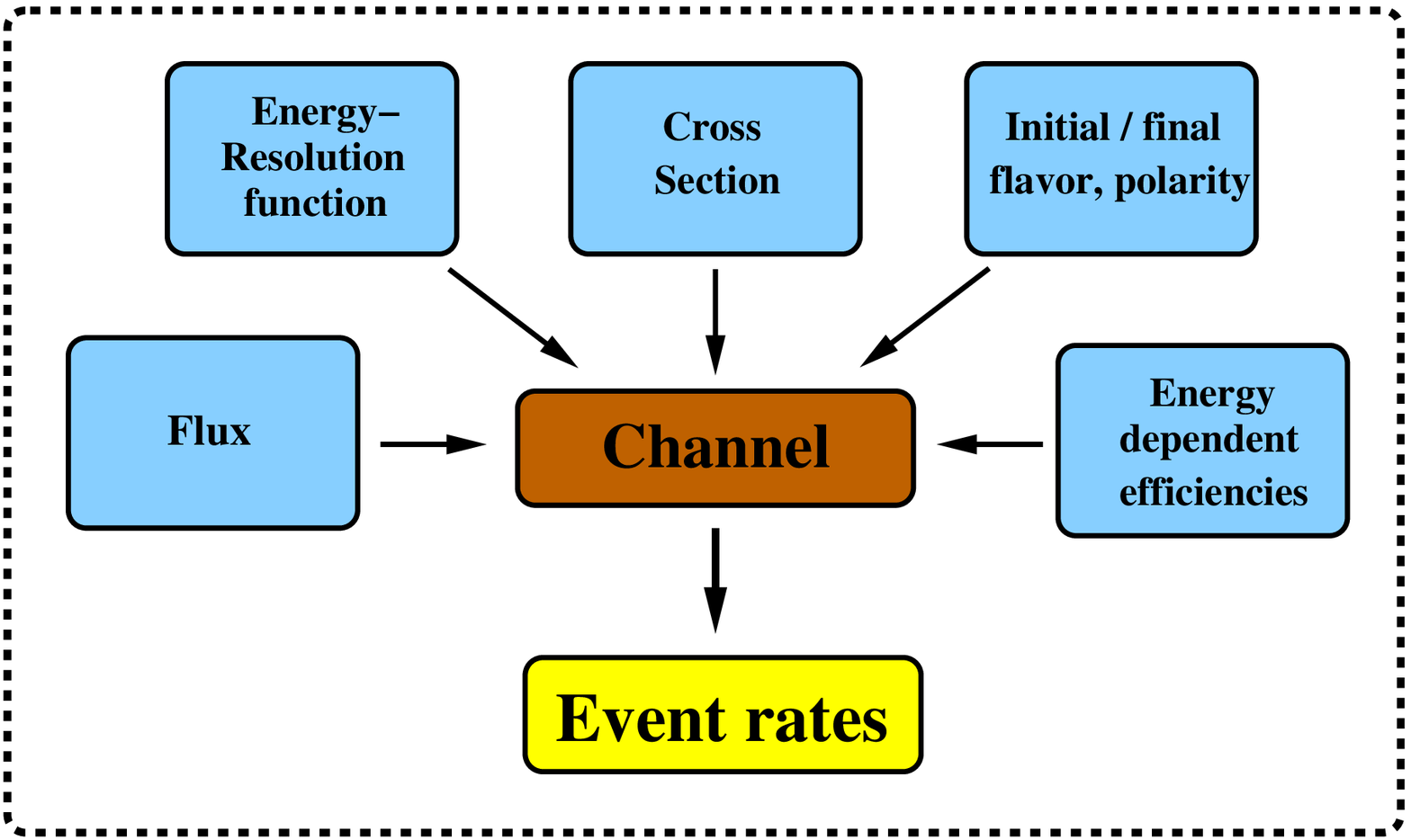} &
\includegraphics[width=4.5cm]{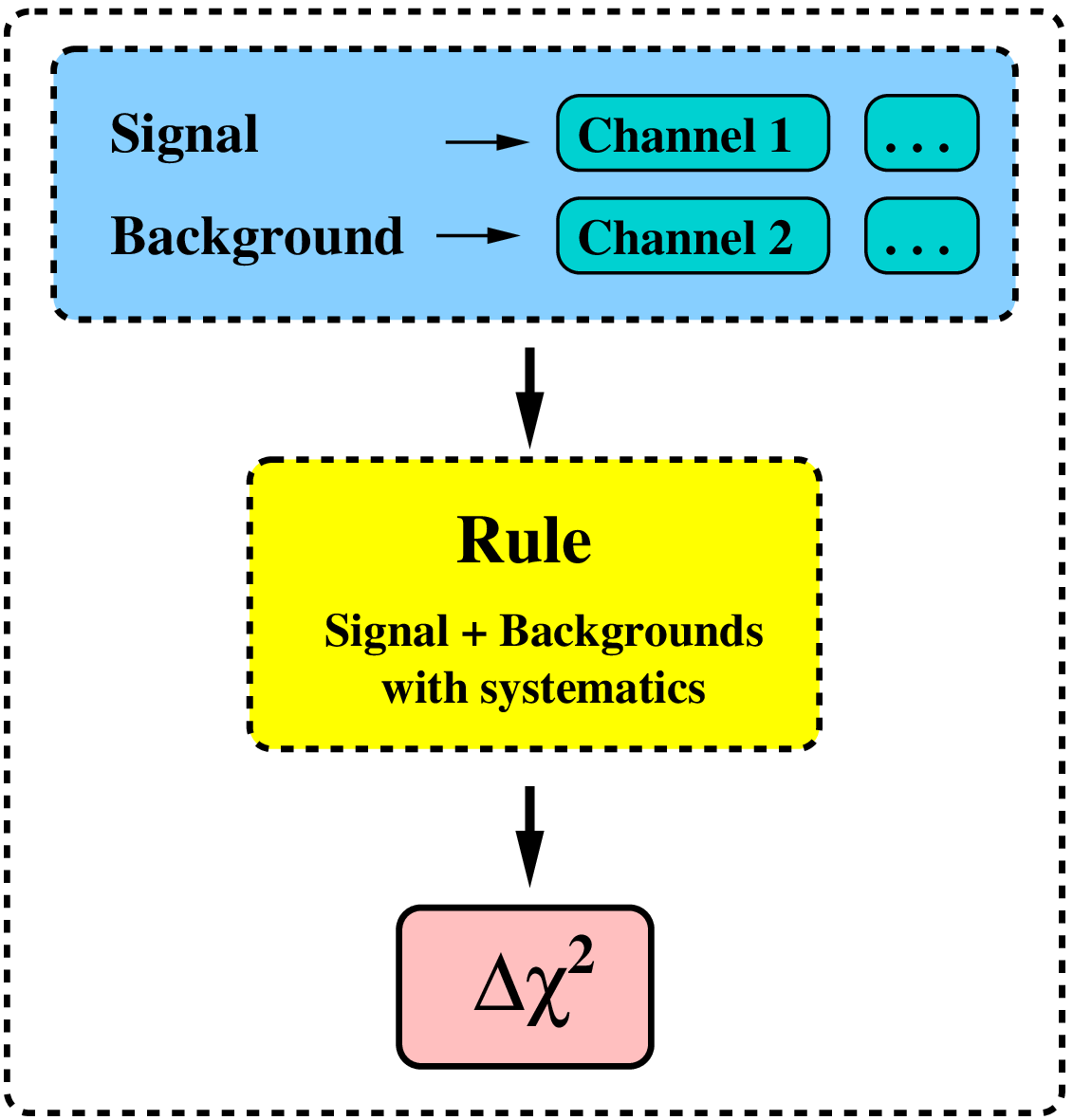} &
\includegraphics[width=4.5cm]{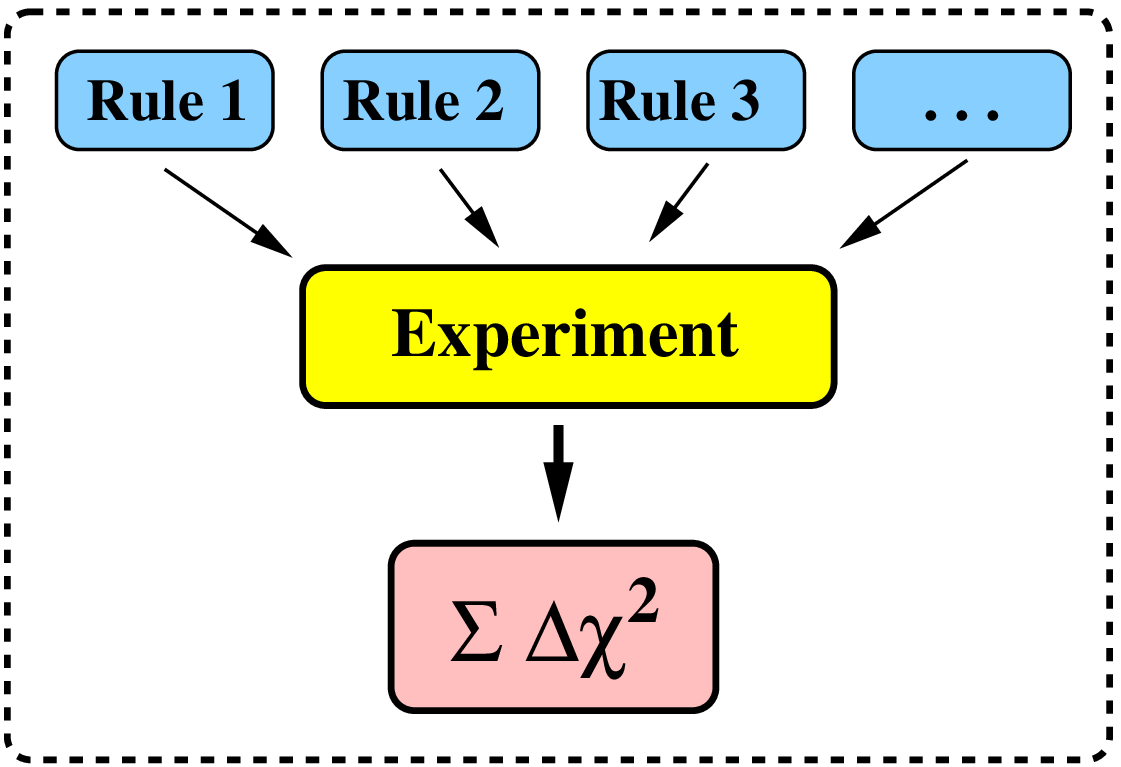} \\
\end{tabular}
\end{center}
\caption{\label{fig:aedl} The most important components of \AEDL :
Channels, rules, and experiments.}
\end{figure}

In order to define experiments, \GLOBES\ uses \AEDL\ (``Abstract
Experiment Definition Language''). An experiment normally corresponds
to an \AEDL\ file, which is a human readable text file written 
in \AEDL\ syntax.

The key concept of \AEDL\ is to regard a detector as an abstract system
which maps the true properties of a neutrino onto the reconstructed properties
of the neutrino. The latter are subsequently used in the fit of the oscillation
parameters. Within \GLOBES , only energy and flavor are  
observables, which is
sufficient for long-baseline and reactor experiments. In  other
experiments, such as atmospheric neutrino experiments, more observables 
(\eg , direction) may have to be considered.

The main components of any \AEDL\ experiment definition are ``channels'',
``rules'', and ``experiments'' (\cf, \figu{aedl}). A channel corresponds to a neutrino oscillation channel including flux, cross section (for one specific interaction type), energy resolution function, initial and final neutrino flavors, their polarity (neutrinos or antineutrinos), and efficiencies. The channel raw event rates are computed according to \equ{events_bin}. Each channel leads to the raw event rates for a 
specific interaction type. In \AEDL , many of the channel components
have to be defined or loaded from files before, such as fluxes, 
cross sections, or the energy resolution function. For the
different options, see the \GLOBES\ manual~\cite{Manual}.

For a ``rule'', the raw event rates of one or more signal channels and
one or more background channels are added. The splitting in signal and
background is arbitrary, but all of the signal and background components
are defined to have a common systematics. The event rates of all 
signal and background components add up to the total event rate of the
rule, which leads to a $(\Delta \chi^2)_r$. The signal or background
within each rule allows the specification of signal and background 
normalization errors and energy tilt or calibration errors. 
These systematical errors are evaluated with the ``pull method'' such as 
in \Refs~\cite{Huber:2002mx,Fogli:2002pt}.  
In addition to these
systematics errors, an overall evaluation strategy is assigned 
to each rule, which specifies the type of systematics (tilt or calibration
error), and the use of spectral information or total event rates.   

Finally, one ore more rules add up to an experiment, where the
total $\Delta \chi^2$ is obtained as the sum of the $(\Delta \chi^2)_r$
of all rules. This approach allows the definition of 
appearance and disappearance channels, neutrino and antineutrino running, 
or interaction types with different systematics (spectral information
versus counting rate) within one experiment. The \GLOBES\ user interface
allows the simulation of one or more experiment simultaneously, which
means that one could also use different experiments for different
oscillation channels. However, there is still one component missing
in the experiment definition, which is the matter density profile.
For an experiment, an arbitrary matter density profile can be specified, 
which is evaluated with the evolution operator method~\cite{Ohlsson:1999um}.
In addition, the matter density profile is multiplied by a scaling factor, 
which is treated as an independent parameter
with a relative (matter density) uncertainty. With this
approach, one can take into account that the matter density profile along a 
specific baseline is only known to about $3\%-5\%$. Thus, for one
particular baseline, all rules should be defined within one experiment.

\section{Analysis of experiments: The C user's interface}

\begin{figure}[t]
\begin{center}
\includegraphics[width=16cm]{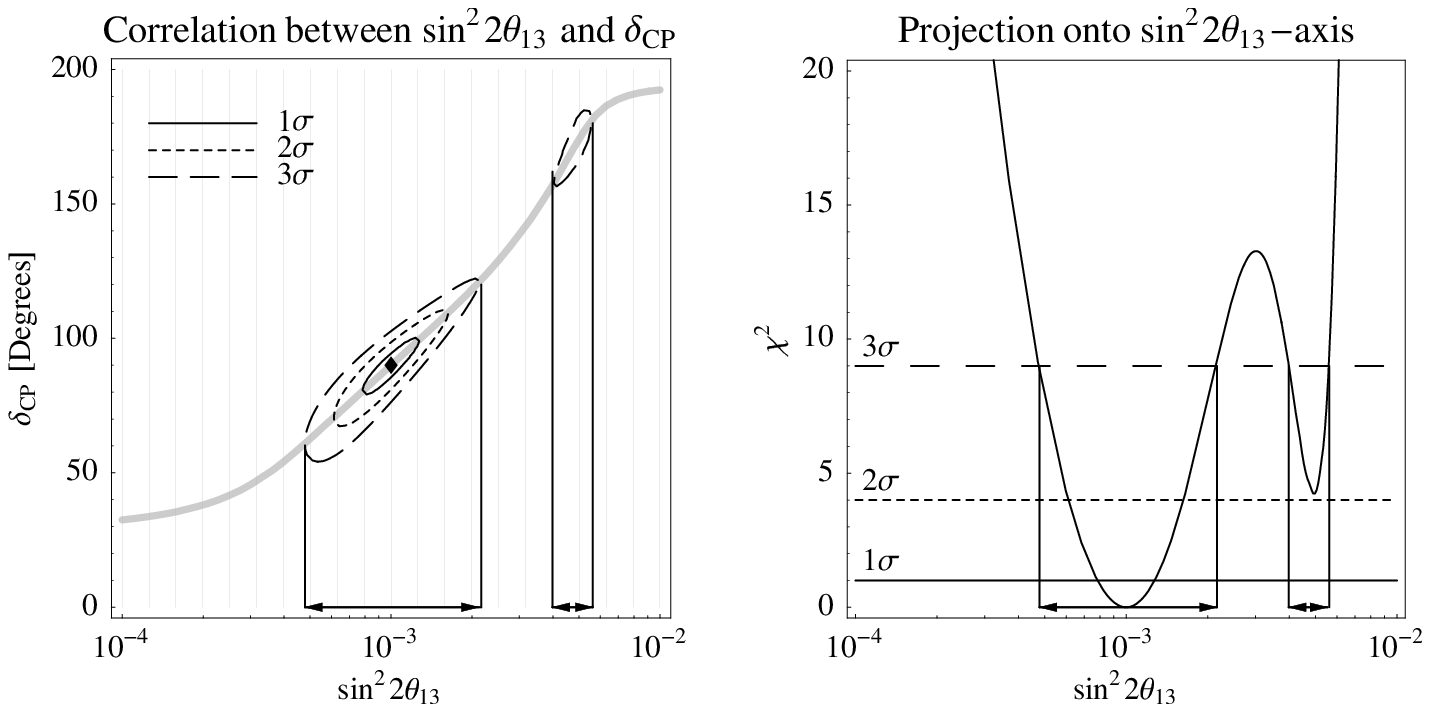}
\end{center}
\mycaption{\label{fig:projex} Left plot: Example for a  correlation between $\stheta$ and $\deltacp$ (for 1 d.o.f. only, un-shown oscillation
parameters fixed). Right plot: The $\chi^2$-value of the projection onto the $\stheta$-axis as function of $\stheta$. The projection onto the  $\stheta$-axis is obtained by finding the minimum $\chi^2$-value for each fixed value of $\stheta$ in the left-hand plot, \ie, along the gray vertical lines. The thick gray curve marks the position of these minima in the left-hand plot. The arrows mark the obtained fit ranges for $\stheta$ at the $3 \sigma$ confidence level (1 d.o.f.), \ie , the precision of $\stheta$.}
\end{figure}

In order to use \GLOBES , a C-library is linked with the application
software. This library provides the user's interface functions.
It allows to load \AEDL\ files, initialize the experiments, and
have access to various $\Delta \chi^2$-functions including any combination
of systematics and correlations. In addition, it provides low-level
access to oscillation probabilities and event rates, and allows
the readout and modification of many experiment parameters at run time.

The most sophisticated feature of the user's interface is the
possibility to include the full multi-parameter correlation among
all oscillation parameters. Together with the matter densities
of the $N_{\mathrm{exp}}$ loaded experiments ($N_{\mathrm{exp}} \ge 1$), the neutrino oscillation parameter space in \GLOBES\ has $6+N_{\mathrm{exp}}$ dimensions. In the $\chi^2$-approach,
the allowed region of $k$ ($1\le k<6+N_{\mathrm{exp}}$) fit parameters (at the chosen confidence level) is obtained as the projection of the $6+N_{\mathrm{exp}}$-dimensional fit manifold onto the $k$-dimensional hyperplane. For example,
for the precision of the parameter $\stheta$ only, the fit manifold
needs to be projected onto the $\stheta$-axis. One can easily imagine
that the computation of the full multi-parameter correlation is
very expensive with a grid-based method. However, the topology
of the neutrino oscillation parameter space is well investigated and
rather smooth. Therefore, it is feasible to use local minimization for each
solution in the parameter space.  For this process, 
a $6+N_{\mathrm{exp}}$-dimensional local minimizer
is started at the ``guessed'' position for the solution
(such as from analytical knowledge) and then runs into a
local minimum. This exercise has to be repeated till all local 
minima, \ie\ degenerate solutions, are found. In \figu{projex},
this minimization is illustrated at the simple example of a 
two-parameter correlation at a neutrino factory.

As we have indicated above, the matter density of
each experiment is treated as an additional parameter, which is known
from external measurement with a certain precision. In principle, 
the \GLOBES\ user's interface allows such externally imposed
precisions for all oscillation parameters. This feature turns out
to be especially useful for input from the solar neutrino experiments,
since they provide better precisions for the solar neutrino
oscillation parameters $\sdm$ and $\theta_{12}$ than long-baseline
 experiments themselves.

\section{Complexity and computational issues}

The \GLOBES\ software only uses polynomial algorithms, which are,
however, suffering from the large number of dimensions.
The $\Delta \chi^2$ computation basically consists of two steps:
The systematics $(\Delta \chi^2)_{\mathrm{sys}}$ computation with {\tt glbChiSys},
and the projection algorithm which uses {\tt glbChiSys} many times. 
The overall computation time is obtained as the product of these two steps.

The run time of the systematics $(\Delta \chi^2)_{\mathrm{sys}}$ 
is 
\begin{equation}
\mathrm{t_{sys}} = \mathcal{O} \left( N_{\mathrm{Sampling}} \times N_{\mathrm{Bins}}  \, + \, N_{\mathrm{Sampling}} \times N_{\mathrm{Layers}} \right) ,
\end{equation}
since the oscillation probabilities for all channels and energies
are only computed once and stored in a list. Here $N_{\mathrm{Sampling}}$ is the number of sampling points for the evaluation of the right integral in \equ{events_bin}, $N_{\mathrm{Bins}}$ is the number of energy bins,
and $N_{\mathrm{Layers}}$ is the number of matter density layers. Thus, whenever the
number of matter density steps is small, the product of sampling points
and bins dominates. In practice, the $N_{\mathrm{Sampling}} \times N_{\mathrm{Bins}}$ energy smearing matrix is already computed at 
the experiment initialization in order to save run time. In addition,
since it contains a lot of zeros, \GLOBES\ uses a special format to
avoid looping over the zeros.

The run time for the projected $\Delta \chi^2$ would, for a grid based method, be
\begin{equation}
 \mathrm{t_{proj}^{Grid}} = \mathcal{O}\left( n^{6+N_{\mathrm{exp}}}\right) \, \times \, \mathrm{t_{sys}},
\end{equation}
where $n$ is the number of grid points to be evaluated for each oscillation
parameter, and $N_{\mathrm{exp}}$ is the number of experiments
(since the matter densities of all experiments appear as parameters). 
Though it is, in principle, possible to use such a method with \GLOBES ,
it is impracticable in most cases, since it would involve at least 
billions of
steps for the complete multi-parameter correlation. The local $6+N_{\mathrm{exp}}$-dimensional minimization reduces this effort
to 
\begin{equation}
\mathrm{t_{proj}^{Minimizer}} = \mathcal{O}(1000) \, \times \,
N_{\mathrm{Deg}} \,  \times \, \mathrm{t_{sys}}
\end{equation}
steps, where $N_{\mathrm{Deg}}$ is the number of (disconnected) 
local minima, \ie, the number of
degeneracies (typically $\mathcal{O}(10)$).

Eventually, \GLOBES\ needs about $10$ to $15$ seconds to compute
a projected $\Delta \chi^2$ on a modern Pentium machine. Therefore,
more sophisticated applications, such as a two dimensional plot
for CP violation as function of the simulated values of $\stheta$ and
$\deltacp$ can be obtained in several hours.

\section{Use of package}

The \GLOBES\ package~\cite{Globes} is a tar-ball which consists of the
source code for the C-library providing the user's interface, the source code
 for a program to test
\AEDL\ files, example C-files illustrating the use of the library, 
experiment prototypes in \AEDL\ with supporting flux and cross 
section files, and the usual 
supporting files for installation and compilation.
Currently, \GLOBES\ is only provided for GNU/Linux.
In addition, an extensive manual covering the user's interface and 
\AEDL~\cite{Manual} 
can be obtained from the web-site. \GLOBES\ is free software and as such 
licensed under the GNU Public License.

The installation of \GLOBES\ is highly automated by the use of the
Autotools family and follows the usual procedures. By default
a shared library {\tt libglobes.so} is installed, but 
a static version is available, too. In addition, an executable named {\tt globes} is installed. It is another possibility to install \GLOBES\ without
root privilege into a user's home directory. During the automatic
configuration process also an example Makefile is produced which can
serve as a template for compiling and linking own applications against
{\tt libglobes.so}. This Makefile can be used to directly compile and
link the examples
from the manual. 

\AEDL\ files, such as the experiment prototypes,
can be edited with any text editor, in order to be loaded by the
users's interface later. In addition, the {\tt globes} binary allows to
develop and test experiment descriptions. For example, event rate
information can be quickly provided to adjust the neutrino flux
normalization. For further information, we refer
at this place to the \GLOBES\ manual~\cite{Manual}.

\section{Summary and conclusions}

In summary, the \GLOBES\ software package provides powerful tools
to do a full three-flavor analysis of future long-baseline and reactor
 experiments
including systematics, correlations, and degeneracies. In addition,
the abstract experiment definition language allows to define
experiments at an abstract level in a highly accurate fashion. 
Some of the major strengths of \GLOBES\ are the ability to quickly
define new experiments, the potential to take into account the
full multi-parameter correlation, the possibilities to include
external input and matter density uncertainties and 
the ability to analyze several experiments simultaneously.

We conclude that the \GLOBES\ software has two major target groups:
Experimentalists, who want to quickly evaluate the physics potential
of their setups, and theorists, who want to test new ideas or
strategies with pre-defined experiments. Especially, the separation
between experiment descriptions as simple text files (together
with supporting files) and the application software should allow an
efficient and lively interaction between those two target groups.

\subsection*{Acknowledgments}

We would like to thank Martin Freund, who wrote the very first
version of a three-flavor matter profile treatment many years ago.
PH is especially thankful for the invaluable advice of Thomas Fischbacher on
many design issues. 
In addition, we would like to thank
Mark Rolinec for his help to translate the experiment descriptions
into \AEDL , and for creating the illustrations in this manuscript.
Finally, thanks to all the people who have been pushing this project
for many years, to the ones who have been continuing asking for the 
publication of the software, and the referees of several of our
papers for suggestions which lead to improvements in the software.
This work has been supported by the ``Sonderforschungsbereich
375 f\"ur Astro-Teilchenphysik der Deutschen Forschungsgemeinschaft''.


\end{document}